\documentclass[aps,prb,showpacs, groupedaddress,twocolumn, superscriptaddress]{revtex4}
\usepackage{amsmath,graphicx,latexsym,times,color}
\usepackage{setspace}
\usepackage{hyperref}
\usepackage{array}
\usepackage{titlesec}
\usepackage{xspace}
\usepackage{appendix}

\begin{document}

\title{Magnetocrystalline anisotropy of Fe and Co slabs and clusters on SrTiO$_3$ by first-principles}

\author{Dongzhe Li}
\affiliation{SPEC, CEA, CNRS, Universit\'e Paris-Saclay, CEA Saclay, F-91191 Gif-sur-Yvette Cedex, France}

\author{Cyrille Barreteau}
\affiliation{Service de Physique de l'Etat Condens\'e, Centre National de la Recherche Scientifique, Unit\'es Mixtes de Recherche 3680, IRAMIS/SPEC, CEA Saclay, Universit\'e Paris-Saclay, F-91191 Gif-sur-Yvette Cedex, France}
\affiliation{DTU NANOTECH, Technical University of Denmark, {\O}rsteds Plads 344, DK-2800 Kgs. Lyngby, Denmark}

\author{Alexander Smogunov}
\email{alexander.smogunov@cea.fr}
\affiliation{Service de Physique de l'Etat Condens\'e, Centre National de la Recherche Scientifique, Unit\'es Mixtes de Recherche 3680, IRAMIS/SPEC, CEA Saclay, Universit\'e Paris-Saclay, F-91191 Gif-sur-Yvette Cedex, France}

\date{\today}

\begin{abstract}
In this work, we present a detailed theoretical investigation of the electronic and magnetic properties of ferromagnetic slabs and clusters deposited on SrTiO$_3$ via first-principles, with a particular emphasis on the magneto-crystalline anisotropy (MCA). We found that in the case of Fe films deposited on SrTiO$_3$ the effect of the interface is to quench the MCA whereas for Cobalt we observe a change of sign of the MCA from in-plane to out-of-plane as compared to the free surface. We also find a strong enhancement of MCA for small clusters upon  deposition on a SrTiO$_3$ substrate. The hybridization between the substrate and the $d$-orbitals of the cluster extending in-plane for Fe and out-of-plane for Co is at the origin of this enhancement of MCA. As a consequence, we predict that the Fe nanocrystals (even rather small) should be magnetically stable and are thus good potential candidates for magnetic storage devices.
\end{abstract}


\newcommand{\Ea}{\ensuremath{E_a}}
\newcommand{\Eb}{\ensuremath{E_b}}
\newcommand{\Ec}{\ensuremath{E_c}}
\newcommand{\Ed}{\ensuremath{E_d}}
\newcommand{\Ee}{\ensuremath{E_e}}

\newcommand{\eV}{\ensuremath{\,eV}}

\maketitle

\section{ Introduction }
\label{Intro}

The fine-tuning of the interfacial magnetocrystalline anisotropy (MCA) in ferromagnet-oxide insulator systems represents a key issue for several technological applications such as perpendicular magnetic tunnel junctions (p-MTJs) \cite{Nistor2009, Kim2008, Mizunuma2009} and tunneling anisotropic magnetoresistive (TAMR) systems \cite{Park2008, Gao2007}. It is well known that the physical origin of the MCA is the spin-orbit coupling (SOC). For the 3$d$ transition-metals the SOC being of the order of a few tens of meV, the MCA per atom is extremely small (10$^{-3}$ meV) in the bulk phase of cubic materials but can get larger ($\sim$ 10$^{-1}$ meV) at surfaces/interfaces   due to reduced symmetry. In order to obtain even larger MCA, traditionally, the MCA of nanostructures of 3$d$ elements is enhanced by introducing 4$d$ or 5$d$ heavy elements with large SOC as a substrate such as Co/Pt \cite{Nakajima1998} or Co/Pd \cite{Weller1994} multilayers as well as in small 3$d$ clusters on heavy elements substrate \cite{Bornemann2012}. However, despite the weak SOC at the interface, a strong MCA has been observed in Co and Fe thin films on metallic oxides such as AlO$_x$ and MgO \cite{Monso2002, Ikeda2010}. The origin of this large MCA is attributed to electronic hybridization between the metal 3$d$ and O-$2p$ orbitals \cite{Yang2011}. More recently, Ran et $al$. have shown that it was possible to reach the magnetic anisotropy limit ($\sim$ 60 meV) of 3$d$ metal atom by coordinating a single Co atom to the O site of an MgO surface \cite{Rau2014science}. Enhancing MCA of nanostructures provides a route towards future miniaturization of data storage at ultimate length scales \cite{Sebastian2012, Khajetoorians2014}

In our previous work, we demonstrated that for both Fe and Co nanocrystals, the MCA of free nanocrystals is mainly dominated by the (001) facets resulting in an opposite behavior: out-of-plane and in-plane magnetization direction favored in Fe and Co nanocrystals, respectively \cite{Dongzhe2013, Dongzhe2014}. Therefore, the study of magnetic properties of nanocrystals deposited on a SrTiO$_3$ as experimentallly obtainable\cite{Silly2005, Dongzhe2014} is essential, since depending on the bonding between the substrate and (001) facets this can influence greatly the overall behaviour of the nanocrystal. In this paper, we report first-principles investigations of the MCA of  bcc-Fe(001) and fcc-Co(001) deposited on a SrTiO$_3$ substrate, namely Fe(Co)$|$SrTiO$_3$ interface. Next, we also investigated the MCA of very small (five atoms)  Fe and Co clusters on SrTiO$_3$.

\section{Calculation method}
\index{DFT_method}

We carried out the first-principles calculations by using the plane wave electronic structure package QUANTUM ESPRESSO (QE) \cite{Giannozzi2009}. Generalized gradient approximation in Perdew, Burke and Ernzerhof parametrization \cite{Perdew1996} was used for electronic exchange-correlation functionals and a plane wave basis set with the cutoffs of 30 Ry and 300 Ry were employed for the wavefunctions and for the charge density, respectively. The MCA was calculated from the band energy difference between two magnetic orientation ${\hat {\bf m}_1}$ and ${\hat {\bf m}_2}$ using force theorem \cite{Dongzhe2014}, as we implemented recently in QE package:
\begin{equation}
\text{MCA}=\sum_{\alpha \text{occ}}\epsilon_{\alpha}({\hat {\bf m}_1})-\sum_{\alpha \text{occ}}\epsilon_i({\hat {\bf m}_2}).
\end{equation}

Where $\epsilon_{\alpha}({\hat {\bf m}})$ are the eigenvalues obtained after a single diagonalization of the Hamiltonian including SOC, but starting  from an  initial  charge/spin density of a self-consistent scalar-relativistic calcutation that has been rotated to the appropriate spin orientation axis as explained in Ref. \onlinecite{Dongzhe2014}.

The Fe(Co)$|$SrTiO$_3$ interface was simulated by 10 layers of bcc-Fe(001)[fcc-Co(001)] slab deposited on a SrTiO$_3$(001) with 5 layers. In the ionic relaxation, the Brillouin-zone has been discretized by using 10 $\times$ 10 in-plane $k$-points mesh and a smearing parameter of 10 mRy. Two bottom layers of SrTiO$_3$ were fixed while other three layers of substrate and ferromagnetic slabs were relaxed until the atomic forces are  less than 1 meV/\AA. To obtain reliable values of MCA, the convergence of calculations has been carefully checked. A mesh of 20 $\times$ 20 in-plane $k$-points has been used for SCF calculation with scalar-relativistic PPs with a smaller smearing parameter of 5 mRy. In non-SCF calculation with full-relativistic PPs including SOC the mesh was increased to 60 $\times$ 60 and smearing parameter was reduced down to 1 mRy which provides an accuracy of MCA below 10$^{-2}$ meV.

For small Fe and Co clusters on SrTiO$_3$, the interface was simulated by a $(4 \times 4)$ in-plane TiO$_2$-terminated SrTiO$_3$(001) substrate with 5 atomic layers containing one Fe(Co) cluster made of 5 atoms. Two bottom layers were fixed while other three layers of substrate and Fe(Co) cluster were relaxed until atomic forces are less than 1 meV/\AA. For both scalar and full relativistic calculations, a $(8 \times 8 \times 1)$ $k$-points mesh and a smearing parameter of 1 mRy was used. In addition, the effect of unphysical interaction in the direction $z$ was minimized by taking a vacuum space of about 15 \AA. 

\section{Results and discussions}
\index{Results}

\subsection{Fe(Co)$|$SrTiO$_3$ interfaces}
\index{slab_STO}

\begin{figure}[htbp]
\begin{center}
\includegraphics[width=1\linewidth]{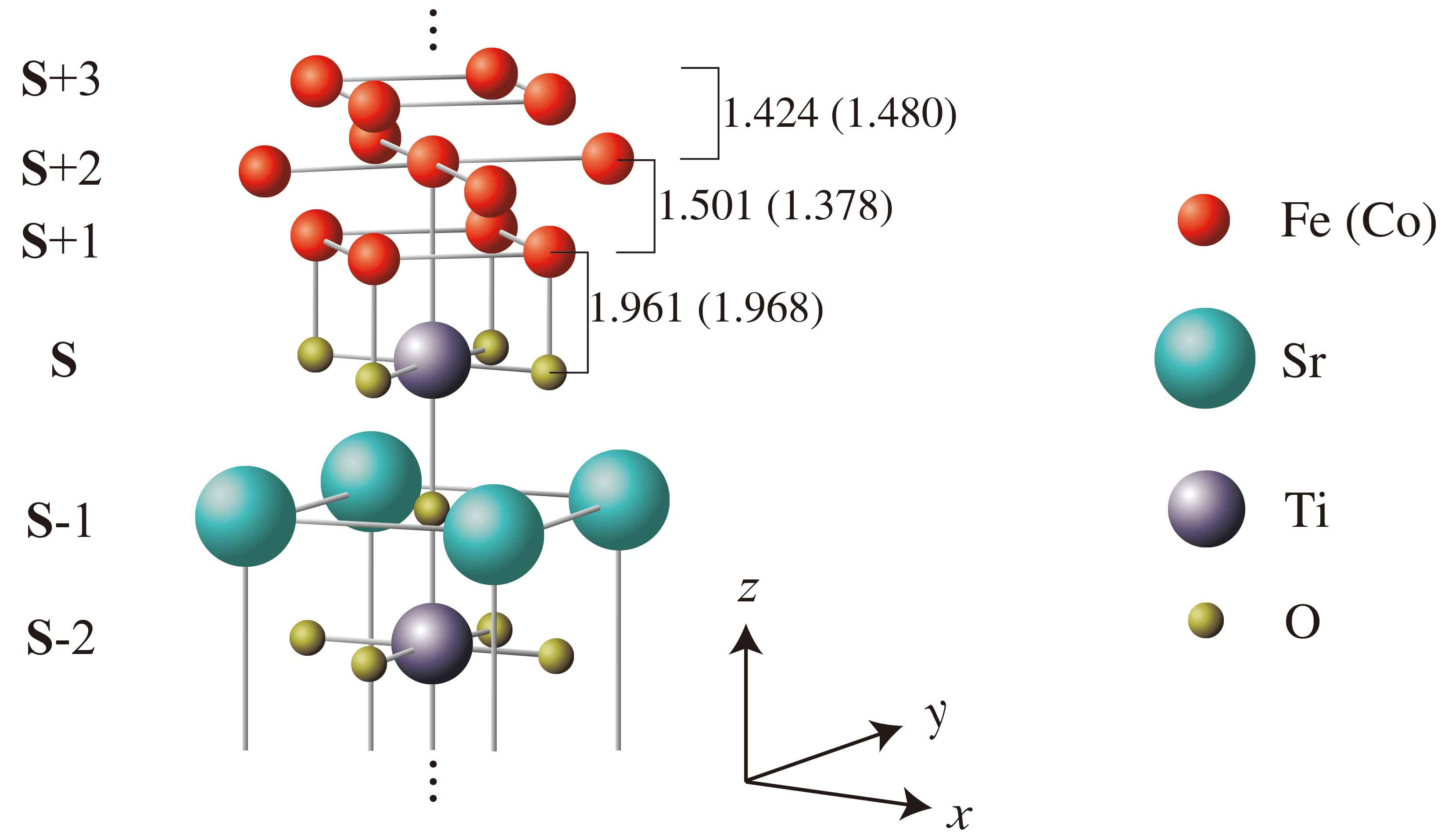}
\caption{Atomic structure of bcc-Fe(001) and fcc-Co(001) slabs on top of TiO$_2$-terminated (001) surface of SrTiO$_3$. The ferromagnetic slab is rotated by 45$^{\circ}$ with respect to substrate in order to better match with the SrTiO$_3$ lattice. Note that each layer of ferromagnetic slabs is made of 2 atoms per supercell. Layers $\bf S+$3, ..., $\bf S-$2 are shown and the distances in the $z$ direction between different layers are also indicated.}
\label{interface_structure}
\end{center}
\end{figure}

The SrO and TiO$_2$ planes in the perovskite cubic SrTiO$_3$ alternate in the (001) direction, here SrTiO$_3$(001) surface was chosen to be TiO$_2$-terminated since it is energetically more favorable than SrO-terminated one \cite{Fechner2008}. The lattice constants of bulk bcc-Fe, fcc-Co and SrTiO$_3$ are 2.85, 3.53 and 3.93 \AA, as compared to the experimental values of 2.87, 3.54 and 3.91 \AA. When deposited on SrTiO$_3$  the in-plane lattice parameter of Fe(Co) slab is imposed by the one of  bulk SrTiO$_3$ since it has been shown that the Co layer can  nicely be grown on this substrate \cite{Teresa1999prl, Teresa1999science}. In order to obtain a better match, the Fe and Co slabs are rotated by 45$^{\circ}$ with respect to the substrate, and each layer of the ferromagnetic slab is made of 2 atoms per supercell. The TiO$_2$ layer at the interface in Fe(Co)$|$SrTiO$_3$ is denoted as $\bf S$ (see Fig. \ref{interface_structure}). Layers toward the SrTiO$_3$ bulk are labeled as $\bf S$$-$1, $\bf S$$-$2, etc., while Fe(Co) layers towards the surface are labeled as $\bf S$$+$1, $\bf S$$+$2, $\bf S$$+$3, etc. 

We found that the most stable configuration is, in all cases, where the Fe(Co) sites in layer  $\bf S+$1 are on top of the O sites in layer $\bf S$ with the distance of 1.961(1.968) \AA.  This is in agreement with previous study in Ref. \onlinecite{Oleinik2001}. We used 12 \AA~of vacuum space in the $z$ direction in order to avoid the unphysical interactions between two adjacent elementary unit cells. The mismach with SrTiO$_3$ was found to be about $-$2.5 and 10.1 \% for Fe and Co, respectively. The Fe and Co slabs have been strained and relaxed to accomodate the lattice structure of the SrTiO$_3$ substrate, respectively. As a result, one finds that the distances beween $\bf S$ and $\bf S+$1 of about 1.501 \AA~and 1.378 \AA~which should be compared with the bulk values of 1.425 \AA~and 1.765 \AA~for Fe and Co, respectively.

\subsubsection{Magnetic spin moment}

\begin{figure}[htbp]
\begin{center}
\includegraphics[width=1\linewidth]{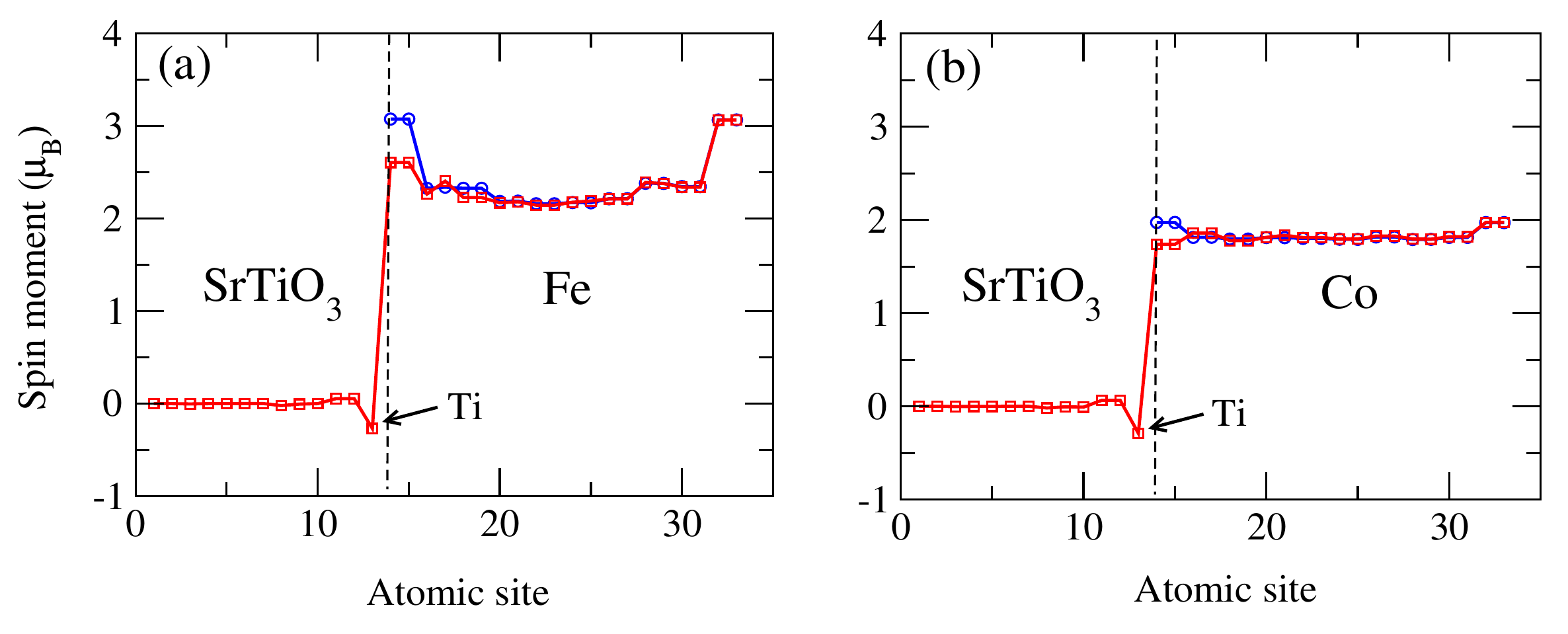}
\caption{Layer-resolved magnetic spin moment (in $\mu_{\text B}$) at Fe$|$SrTiO$_3$(001) (a) and Co$|$SrTiO$_3$(001) (b) interfaces. Blue circles and red squares correspond to free slab and slab on SrTiO$_3$ substrate, respectively.}
\label{slabs_STO_moment}
\end{center}
\end{figure}

We plot in Fig. \ref{slabs_STO_moment} the local  spin moments of a free Fe(Co) slab (blue circles) but for which the ionic positions are the one obtained after relaxation in presence of the SrTiO$_3$(001) substrate. In this way we can evaluate the role of the relaxation on the free surface as compared to the interface. The local spin moments of the full system  Fe(Co)$|$SrTiO$_3$(001)  are shown in red squares. For free slabs, the magnetic moment of $\bf S+$1 layer are enhanced up to 3.07 and 1.97 $\mu_{\text B}$ with respect to their bulk values of 2.15 and 1.79 $\mu_{\text B}$ in $\bf S+$5 layer for Fe and Co, respectively. However, in the case of Fe(Co)$|$SrTiO$_3$, the surface spin moment is reduced to 2.61 and 1.74 $\mu_{\text B}$ (it is even smaller than its bulk value) due to  bonding and charge transfer at the interface. In addition, the hybridization between Fe 3$d$ and states of TiO$_2$ at the interface induces spin moments on Ti and O atoms. It has been found that the induced magnetic moment of the interface O atom in $\bf S$ layer is $\sim$ 0.05 (0.06) $\mu_{\text B}$ and is parallel to the magnetic moment of Fe(Co). A much larger induced but opposite spin moment  in $\bf S$ layer has been found on Ti atoms  :$\sim$ $-$0.27 ($-$0.29) $\mu_{\text B}$. 

\subsubsection{Electronic properties}

In order to explain the origin of the induced magnetic moments at the interface, we investigated the electronic structure (PDOS)  of the free Fe(Co) slab as well as the Fe(Co)$|$SrTiO$_3$ interface compared to the corresponding PDOS in bulk phase of bcc-Fe (fcc-Co) and SrTiO$_3$. 

\begin{figure}[htbp]
\begin{center}
\includegraphics[width=1.0\linewidth]{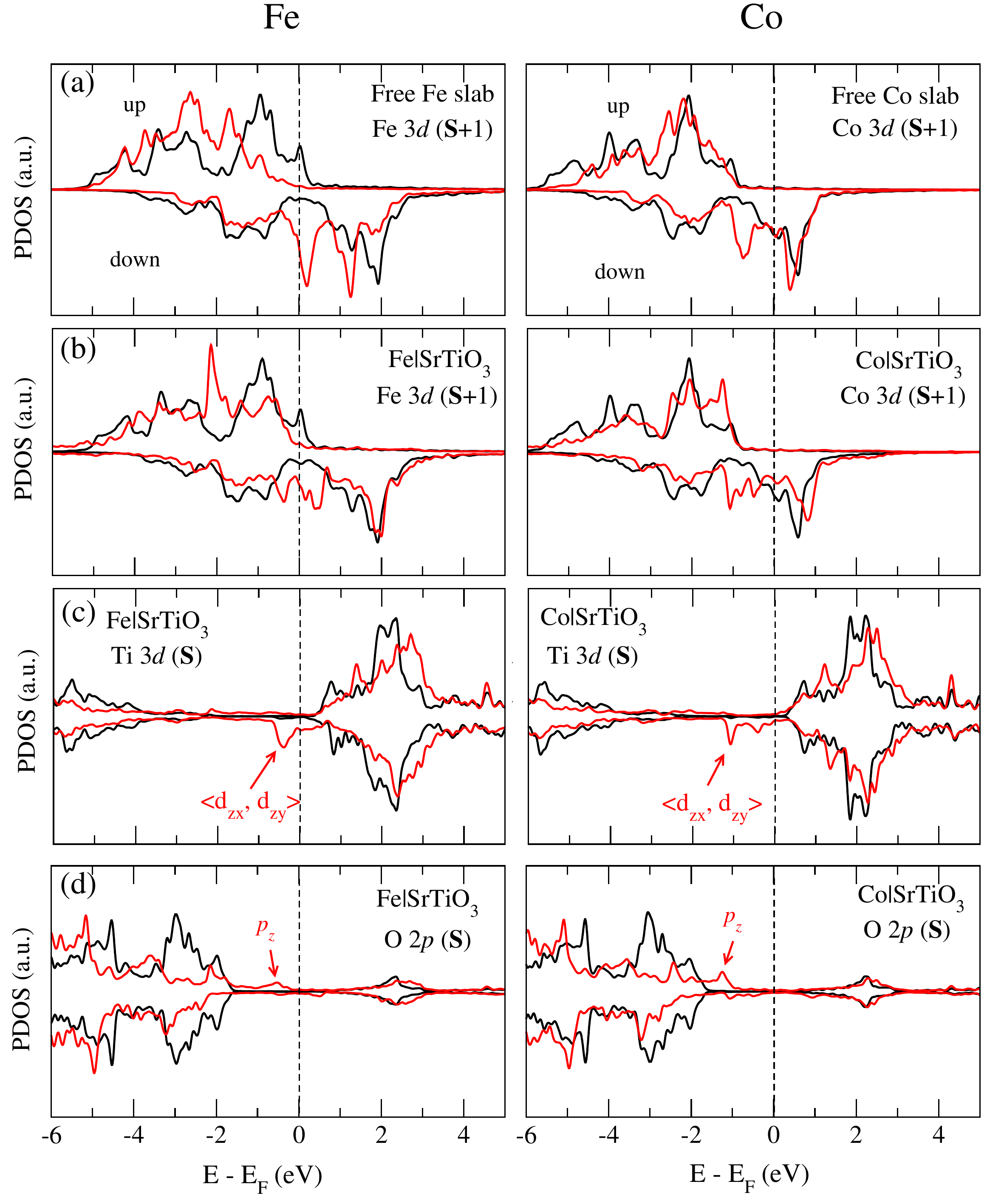}
\caption{(a) Free Fe slab: Scalar-relativistic projected density of states (PDOS) of the surface Fe 3$d$ orbitals in layer $\bf S+$1; b) Fe$|$SrTiO$_3$(001): PDOS of Fe 3$d$ orbitals in layer $\bf S+$1, (c) Ti 3$d$ and (d) O 2$p$ orbitals in layer $\bf S$. The DOS of atoms in the central monolayer of Fe slab (a, b) or (c, d) TiO$_2$ in layer $\bf S-$2 are plotted as black lines. Positive and negative PDOS are for spin up and spin down channels, respectively. The vertical dashed lines indicate the Fermi level ($E_{\text{F}}$). It is the same for Co as presented in the right panels.}
\label{electronic_structure_slabs}
\end{center}
\end{figure}

As shown in Fig. \ref{electronic_structure_slabs} (a), the DOS of the interfacial Fe(Co) 3$d$ ($\bf S+$1) (red line) for free slab differs from the DOS of the bulk Fe(Co) 3$d$ ($\bf S+$5) (black line) as a result of the reduced coordination. A significant minority spin states at $\sim$ 0.1 and 0.7 eV ($-$0.4 and 0.2 eV) with respect to the Fermi level has  been found for Fe(Co) at the interface.These states are the origin of the increase of spin moment for the surface atom as shown in Fig. \ref{slabs_STO_moment}.

Fig. \ref{electronic_structure_slabs} (b) - (d) show the PDOS of Fe(Co) 3$d$ ($\bf S+$1), Ti 3$d$ ($\bf S$) and O 2$p$ ($\bf S$) orbitals at Fe(Co)$|$SrTiO$_3$ interface, indicating the presence of hybridizations between the orbitals. It is well known that the degree of hybridization at the interface depends on the strength of the orbital overlap and inversely on the energy seperation between them. Although there is a direct atomic bonding between the interfacial Fe(Co) and O atoms, the induced magnetic moment on the O atom was found to be relatively small ($\sim$ 0.05 $\mu_{\text B}$). This is due to the fact that O 2$p$ ($\bf S$) orbitals lie well below the Fermi level and, therefore, have a small overlap with the Fe(Co) 3$d$ states. However, the Ti 3$d$ orbitals that are centered at about 2 eV above the Fermi level [the black lines in Fig. \ref{electronic_structure_slabs} (c)] have a strong hybridization with the minority-spin Fe(Co) 3$d$ orbitals which have a significant weight at these energies [the black lines in Fig. \ref{electronic_structure_slabs} (b)]. The most important consequence of this hybridization is the formation of the hybridized  states in the interval of energies [ $-$0.5, $+$0.5 ] eV and [ $-$1, $+$1 ] eV for Fe and Co, respectively. As shown in Fig. \ref{electronic_structure_slabs} (c), the DOS of the Ti 3$d$ $\bf S$ layer at the Fe(Co)$|$SrTiO$_3$ interface, the minority-spin states which originates from the $d_{zx}$ and $d_{zy}$ orbitals at $\sim$ $-$0.5 eV (the two peaks at $-$1 eV and $-$0.5 eV) are occupied, whereas the corresponding majority-spin states are found at $\sim$ $+$1.5 eV (the two peaks at $+$0.5 eV and $+$1 eV) are unoccupied. This leads to an induced magnetic moment of $-$0.27 and $-$0.29 $\mu_{\text B}$ on the Ti ($\bf S$) for Fe and Co based interfaces, respectively.

\subsubsection{Local analysis of MCA}

We now investigate the MCA of the Fe(Co)$|$SrTiO$_3$ interface. The MCA is calculated as band energy difference between the spin quantization axes perpendicular and parallel to the slab surface, explicitely, 
\begin{math}
\text{MCA}=E_{\perp}^{\text{band}}-E_{\parallel}^{\text{band}}
\end{math}, and for the sake of simplicity we have chosen the most symmetric in plane orientation. By definition a positive (negative) sign in MCA means in-plane (out-of-plane) magnetization axis. It should be noted that, the full relativistic Hamiltonian including spin-orbit coupling is given in a basis of total angular momentum eigenstates $|j, m_j>$ with $j=l \pm \frac{1}{2}$. Although the ($l,~m_l,~m_s$) is not a well defined quantum number for the full relativistic calculations, the MCA can still be projected into different orbital and spin by using local density of states. Since the spin-orbit coupling in 3$d$-electron systems is relatively small, this approximate decomposition introduces a negligible numerical inaccuracy.

\begin{figure*}[htbp]
\begin{center}
\includegraphics[width=0.85\linewidth]{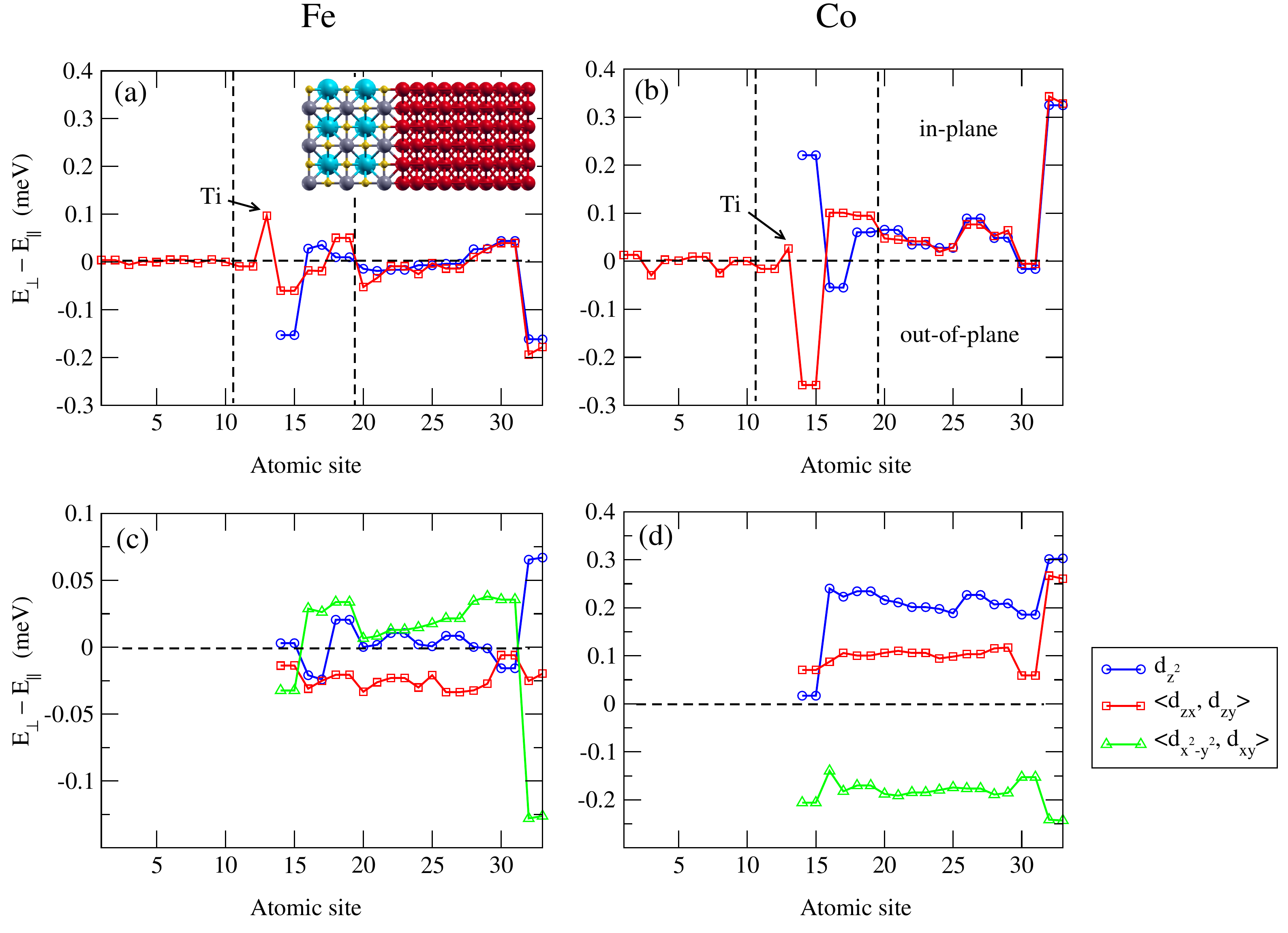}
\caption{Atom-resolved MCA at Fe$|$SrTiO$_3$ (a) and Co$|$SrTiO$_3$ (b) interfaces, blue circles and red squares correspond to free slab and slab on a SrTiO$_3$ substrate, respectively. $d$-orbitals-resolved MCA for Fe (c) and Co (d) slabs on SrTiO$_3$, we plot only the part of ferromagnetic slabs. Due to symmetry, contributions from different orbitals in ($d_{zx}, d_{zy}$) and ($d_{x^2-y^2}, d_{xy}$) pairs are very similar so that their averaged values are presented for simplicity. Note that positive and negative MCA represent in-plane and out-of-plane magnetization, respectively.}
\label{slabsSTO_mca}
\end{center}
\end{figure*}

As shown in Fig. \ref{slabsSTO_mca} (a) and (b), we have calculated the atom-resolved MCA of the Fe(Co)$|$SrTiO$_3$ system (red squares) and compared it with the free Fe(Co) slab (blue circles) containing 10 atomic layers (but relaxed in  presence of the substrate). For free Fe(Co) slab, the total MCA reaches $\sim$ $-$0.49 (1.60) meV per unit-cell favouring an out-of-plane (in-plane) axis of magnetization. If the Fe(Co) slab is in contact with SrTiO$_3$ substrate, the axis of magnetization is preserved but the total MCA is reduced to $\sim$ $-$0.38 (1.02) meV.

From the atom-resolved MCA, one finds that the MCA curves for free slabs are not symmetrical, particularly pronouced for Co, due to (asymmetrical) relaxation effect. The main contribution to MCA is located in the vicinity of the interface, from $\bf S$ layer to $\bf S+$3 layer, marked as vertical dotted line in Fig. \ref{slabsSTO_mca} (a) and (b), and it converges to the expected bulk value in the center of the slab ($\bf S+$5 layer). Interestingly, at the interface, in comparison with free Fe(Co) slab it appears that the contact with SrTiO$_3$ strongly favors in-plane and out-of-plane for Fe and Co, respectively.

For Fe($\bf S+$1), upon adsorption on SrTiO$_3$, the MCA decreases from $\sim$ $-$0.15 to $\sim$ $-$0.06 meV/atom and the out-of-plane magnetization remains. However, in the case of Co($\bf S+$1), the MCA abruply changes from $\sim$ 0.22 to $\sim$ $-$0.25 meV/atom exhibiting magnetization reversal from in-plane to out-of-plane at the same time. For $\bf S+$2 layer, we find a sign change of MCA between free slab and slab on SrTiO$_3$ for both elements, with the MCA difference of $\sim$ 0.04 meV/atom and $\sim$ 0.15 meV/atom for Fe and Co, respectively. For $\bf S+$3 layer, the MCA enhances slightly ($\sim$ 0.05 meV/atom) in-plane MCA when depositing slabs on SrTiO$_3$ for both elements. Furthermore, the Ti atom in $\bf S$ layer [indicated by arrows in Fig. \ref{slabsSTO_mca} (a) and (b)] presents a rather large in-plane MCA of $\sim$ 0.1 meV/atom and a much smaller in-plane MCA of $\sim$ 0.03 meV/atom for Fe and Co-based interfaces, respectively. As a result, for free slabs, the MCA values from $\bf S+$1 layer to $\bf S+$3 layer sum up to the total value of $\sim$ $-$0.22 meV (out-of-plane) and 0.45 meV (in-plane) for Fe and Co. However, when the slabs are supported on SrTiO$_3$, the overall out-of-plane MCA in the vicinity of the surface (here, the $\bf S$ layer is also taken into account) is almost quenched for Fe by $\sim$ 0 meV, and in the case of Co, a spin transition from in-plane to out-of-plane magnetization has been found with a MCA value of $\sim$ $-$0.10 meV.

In order to understand the origin of this difference in MCA between free Fe(Co) slab and Fe(Co)$|$SrTiO$_3$ system, we investigated the $d$-orbitals-resolved MCA of the Fe(Co) atom as shown in Fig. \ref{slabsSTO_mca} (c) and (d). Here, due to symmetry, the contributions to MCA from ($d_{zx}, d_{zy}$)  and ($d_{x^2-y^2}, d_{xy}$) pairs are almost equal, therefore, their averaged values are presented for simplicity. 

In the case of Fe, we notice that going from the free Fe slab to the Fe$|$SrTiO$_3$ system, the MCA of the $d_{z^2}$ (in-plane magnetization) and ($d_{x^2-y^2}, d_{xy}$) (out-of-plane magnetization) orbitals decreases in magnitude, while the MCA of ($d_{zx}, d_{zy}$) orbitals is almost not affected. In addition, quantitatively, the reduction of  MCA  is larger for ($d_{x^2-y^2}, d_{xy}$) than for $d_{z^2}$ due to stronger hybridization between (Fe-$d_{x^2-y^2,~xy}$, Ti-$d_{zx,~zy}$) orbitals than between (Fe-$d_{z^2}$, O-$p_z$) orbitals. This is attributed to the fact that, shown in Fig. \ref{electronic_structure_slabs}, close to the Fermi level, the shape of the electron density for O and Ti suggest that this density has a $p_z$ character and $d_{zx}$ ($d_{zy}$) character, respectively. Moreover, the strong in-plane MCA in Ti ($\bf S$) layer originates from the Ti-$d_{zx,~zy}$ orbitals since there is a significant weight close to Fermi level of minority-spin (Ti-$d_{zx,~zy}$) orbitals [see Fig. \ref{electronic_structure_slabs} (c) left panel]. As a result, the MCA at the interface appears to almost quench the out-of-plane magnetization when the Fe slab is deposited on SrTiO$_3$. Moreover, if we sum over the contribution of the first three layers of Fe slab at the interface, we found that $d_{zx}, d_{zy}$ orbitals tend to maintain the out-of-plane MCA while $d_{x^2-y^2,~xy}$ orbitals tend to favor the in-plane MCA. A similar result has also been reported in Ref. \cite{Hallal2013} in Fe$|$MgO magnetic tunnel junctions. 

In the case of Co, we find that the hybridization between $p_z$ orbitals of O and $d_{z^2}$ (and, to a slightly lesser extent with $d_{zx,~zy}$) of Co plays a crucial role to decrease in-plane MCA  of the free Co slab. On the other hand, the MCA from in-plane ($d_{x^2-y^2,~xy}$) orbitals of Co is less affected due to rather small minority-spin states of (Ti-$d_{zx,~zy}$) close to the Fermi level [see Fig. \ref{electronic_structure_slabs} (c) right panel]. This leads to induce an inverse spin orientation transition from in-plane to out-of-plane in Co$|$SrTiO$_3$ system. A similar result has also been reported in Ref. \cite{Bairagi2015} at C$_{60}$$|$Co interface. 

\subsection{Fe and Co clusters on SrTiO$_3$}
\index{cluster_STO}

\begin{figure}[htbp]
\begin{center}
\includegraphics[width=1.0\linewidth]{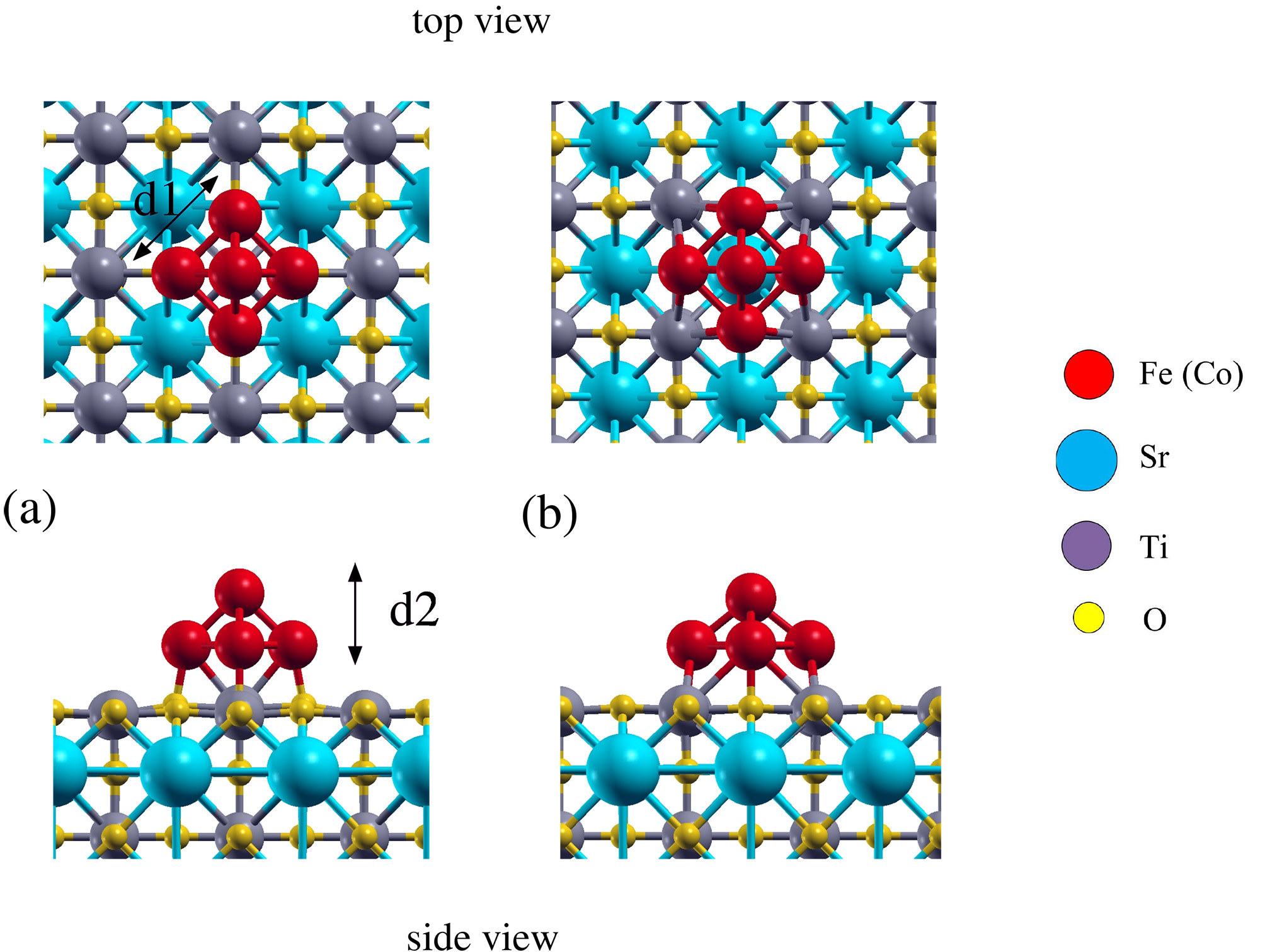}
\caption{Top (upper panels) and side (lower panels) views of the optimized geometries of Fe and Co cluster absorbed on TiO$_2$-terminated SrTiO$_3$(001). Two different adsorption configurations are presented in (a) and (b), the latter one is the most stable configuration for both Fe and Co clusters. The bond length $d_1$ between base atoms and the vertical distance $d_2$ between base and top atoms are indicated.}
\label{pyramid_structure}
\end{center}
\end{figure}

We now investigate the electronic and magnetic properties of Fe and Co clusters deposited on SrTiO$_{3}$ surface. As shown in Fig. \ref{pyramid_structure}, two geometries are examined, namely top (a) and hollow (b) adsorption sites. The base atoms of Fe(Co) clusters are always on top of O atom  for both geometries however  the apex atom  is either on top of a Ti atom (top geometry) or of an underneath Sr atom (hollow geometry). We found that a hollow adsorption site is more energetically stable for both elements, with an energy difference of $\sim$ 0.65 eV and $\sim$ 0.88 eV for Fe and Co, respectively. In the following, we concentrate on the lowest energy configuration.

The strength of the cluster-SrTiO$_3$ interaction can be quantified by calculating the binding energy via the energy difference:
\begin{equation}
E_\text{b}=E[{\text{cluster}}]+E[\text{SrTiO$_3$}]-E[{\text{cluster}|\text{SrTiO$_3$}}]
\end{equation}
where E[{\text{cluster}}], E[\text{SrTiO$_3$}] and E[{\text{cluster}$|$\text{SrTiO$_3$}}] are the total energy of the free cluster, the free SrTiO$_3$ substrate and the cluster-SrTiO$_3$ system, respectively. The calculated binding energy was found to be $\sim$ 4.23 (4.58) eV for Fe(Co) cluster on SrTiO$_3$ substrate, showing strong chemisorption mechanism (see Tab. \ref{table_mag_moment}). 

\begin{center}
\begin{table}[htbp]
\scalebox{0.75}{
 \begin{tabular}{c|cc|ccc}
\hline\hline
\multicolumn{1}{c|}{} & \multicolumn{2}{c|}{Fe} & \multicolumn{2}{c}{Co} \\
\multicolumn{1}{c|}{$$} & \multicolumn{1}{c} {~Free cluster~} & \multicolumn{1}{c|} {~Cluster on SrTiO$_3$~} &\multicolumn{1}{c} {~Free cluster~} & \multicolumn{1}{c} {~Cluster on SrTiO$_3$~}  \\ \hline
	$E_{\text b}$  (eV) & --- & 4.23 & --- & 4.58 \\ 
          $d_1$  (\AA) & 2.31 & 2.55 & 2.17 & 2.20 \\ 
	$d_2$  (\AA) & 1.73 & 1.45 & 1.80 & 1.74 \\ 
	$M_{\text s}^{\text {tot}}$ ($\mu_{\text{B}}$) & 18.00 & 16.63 & 13.00 & 7.67  \\ 
	$|M_{\text s}^{\text {tot}}|$ ($\mu_{\text{B}}$) & 18.34 & 17.96 & 13.41  & 11.06 \\ 
    	$M_{\text s}^{\text {base}}$ ($\mu_{\text{B}}$)  & 3.62 & 3.33 & 2.54 & 1.75  \\
    	$M_{\text s}^{\text {top}}$ ($\mu_{\text{B}}$) & 3.58 & 3.32 & 2.84  & 1.57 \\ \hline\hline
\end{tabular}}
\caption{Binding energies ($E_{\text b}$), atomic bonds, total/total absolute spin moments ($M_{\text s}^{\text {tot}}$/$|M_{\text s}^{\text {tot}}|$), spin moment of base ($M_{\text s}^{\text {base}}$) and top ($M_{\text s}^{\text {top}}$) atoms of the free clusters and clusters deposited on SrTiO$_3$ for the lowest energy configuration.}
\label{table_mag_moment}
\end{table}
\end{center}

Compared to free Fe cluster, the Fe-Fe distance in basal plane ($d_1$) is elongated from 2.31 \AA~to 2.55 \AA~while the Fe-Fe distance in vertical distance from apex to basal plane ($d_2$) is compressed from 1.73 \AA~to 1.45 \AA~ (see Tab. \ref{table_mag_moment}). However, in the case of Co, the geometry optimization of Co$_5$$|$SrTiO$_3$ results in a rather small (negligible) distortion compared to its free Co$_5$ cluster. In addition, the  bond length between Fe(Co) and O is $\sim$ 2 \AA. 

\subsubsection{Magnetic spin moment}

\begin{figure}[htbp]
\begin{center}
\includegraphics[width=1.0\linewidth]{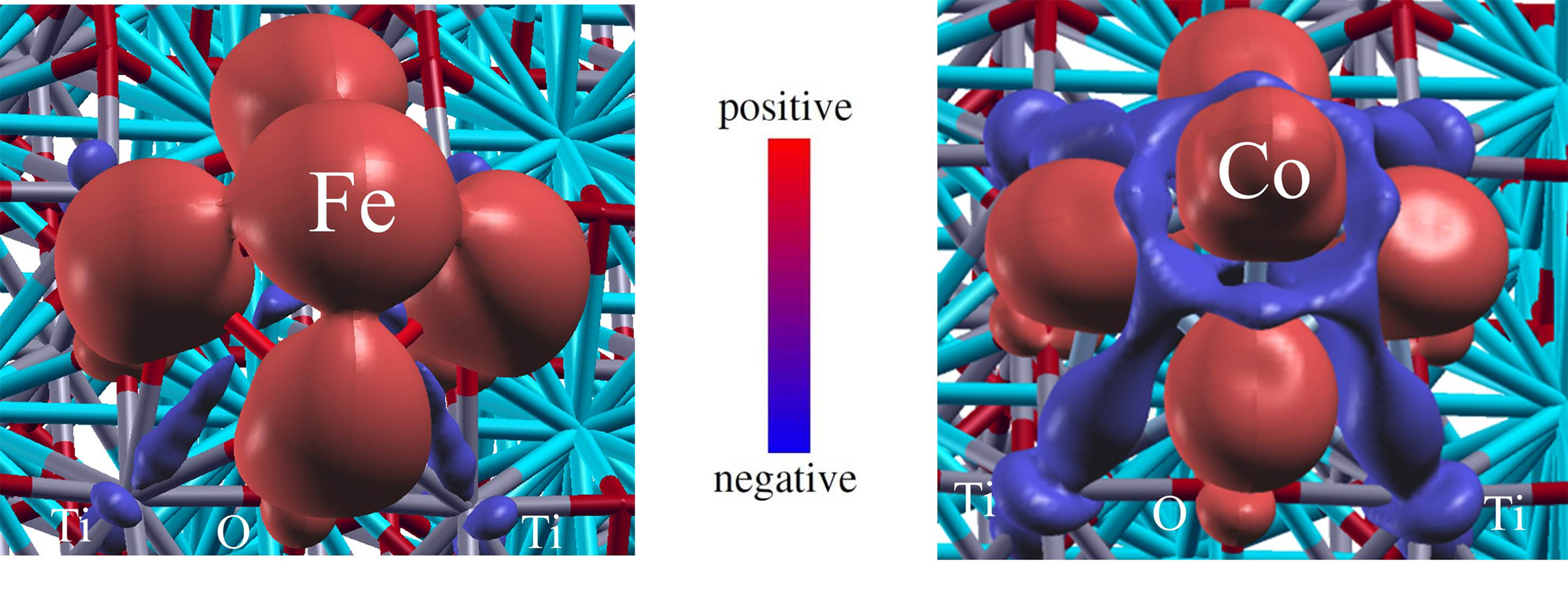}
\caption{Real-space distribution of magnetic spin moment of Fe (left) and Co (right) cluster on SrTiO$_3$. Note that red (blue) corresponds to positive (negative) spin moment. The nonnegligible negative part of spin moment has been found around the Ti atoms at the interface and the Co top atom of cluster.}
\label{Mag_nega_Co}
\end{center}
\end{figure}

We next investigated the local magnetic spin moment. In Tab. \ref{table_mag_moment}, the local spin moments for both free clusters and the clusters on SrTiO$_3$ are given. The binding between Fe(Co) and O atoms reduces the total spin moment from 18.00 $\mu_{\text{B}}$ (free Fe$_5$) to 16.63 $\mu_{\text{B}}$ and from 13.00 $\mu_{\text{B}}$ (free Co$_5$) to 7.67 $\mu_{\text{B}}$ for the deposited clusters. We also calculated the absolute total spin moment $|M_{\text s}^{\text {tot}}|$and compared to corresponding total spin moment $M_{\text s}^{\text {tot}}$. Interestingly, a substantial difference of $\sim$ 3.4 $\mu_{\text{B}}$ has been found between $|M_{\text s}^{\text {tot}}|$ and $M_{\text s}^{\text {tot}}$ for Co$_5$$|$SrTiO$_3$. In order to understand the origin of this difference, we plot in Fig. \ref{Mag_nega_Co} the real-space distribution of magnetic spin moment of Co cluster on SrTiO$_3$. Note that the red (blue) corresponds to positive (negative) spin moment. We can see clearly the negative magnetic moment is mainly localized on Ti atoms at the interface and around the Co top atom of cluster. However, for Fe cluster, the positive spin moment is very localized on the Fe atoms and the negative part is negligible.

\subsubsection{Electronic structure properties}

\begin{figure}[htbp]
\begin{center}
\includegraphics[width=1\linewidth]{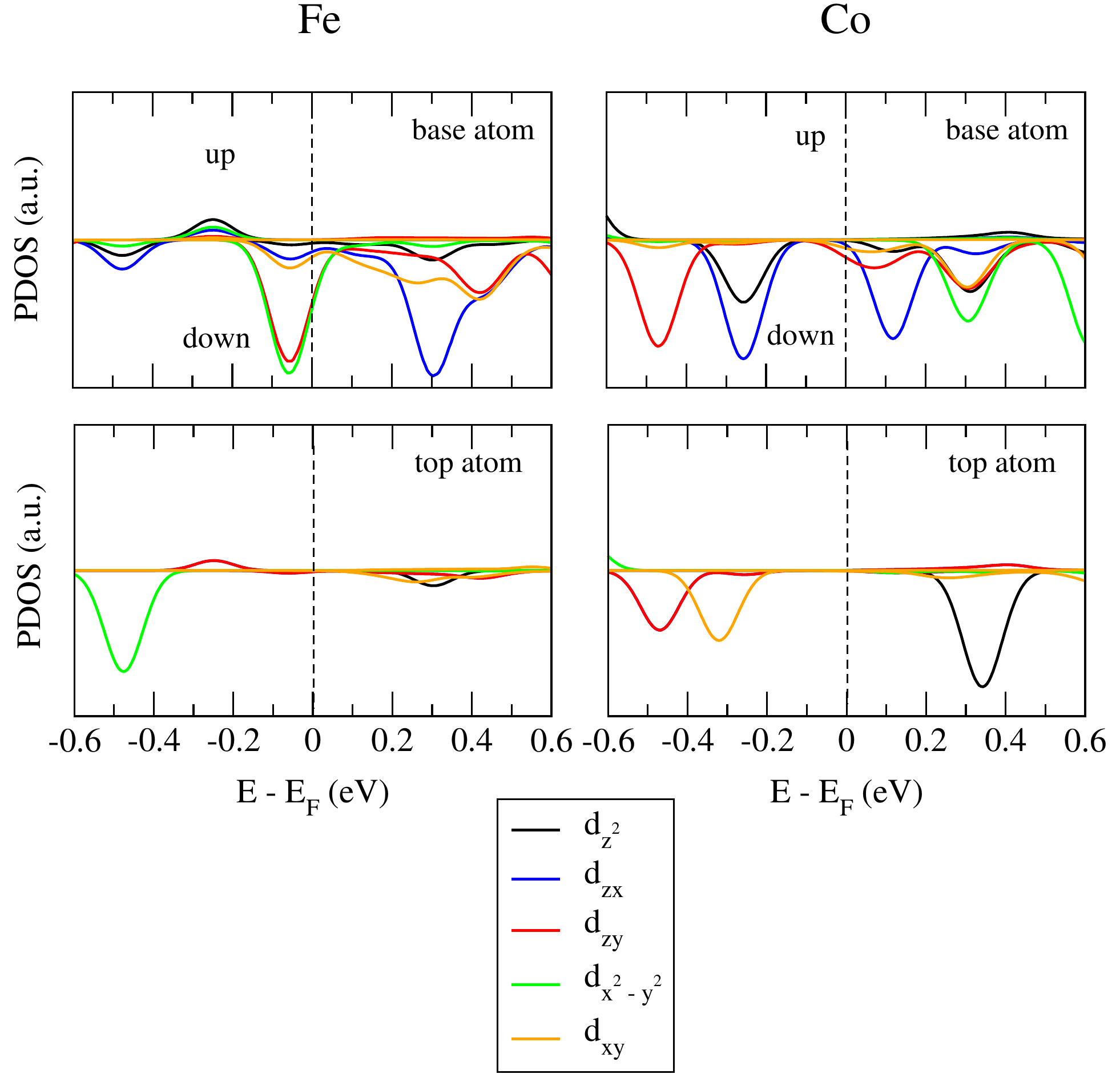}
\caption{Scalar-relativistic $d$-orbitals projected density of states (PDOS) for Fe(Co) base atom (a) and top atom (b) of the cluster absorbed on SrTiO$_3$. Positive and negative PDOS are for spin up and spin down channels, respectively. The vertical dashed lines mark the Fermi level ($E_{\text{F}}$)}
\label{Fe_Co_STO_DOS}
\end{center}
\end{figure}

To gain more insight into the electronic structure of Fe$_5$$|$SrTiO$_3$ and Co$_5$$|$SrTiO$_3$, we plot the scalar-relativistic projected density of states (PDOS) on $d$-orbitals of Fe(Co) base atom and top atom of the cluster in Fig. \ref{Fe_Co_STO_DOS} (a) and (b). 

For the base atom of both clusters, the density of majority states is almost completely occupied (situated below $-0.6$ eV) and negligibly small around the Fermi level, while the density of minority states is partially occupied. Around the Fermi level, there is a higher density of ($d_{x^2-y^2}, d_{xy}$, $d_{zy}$) states for Fe while the most dominant states are the out-of-plane $d$ orbitals for Co, namely ($d_{z^2}, d_{zx}, d_{zy}$) orbitals. For top atom, in the interval of energies [$-$0.25, $+$0.25] eV, the density of states for both majority and minority spins is negligibly small for both clusters.

\subsubsection{Local analysis of MCA}
\label{sec:local-MCA}
\begin{figure*}[htbp]
\begin{center}
\includegraphics[width=0.85\linewidth]{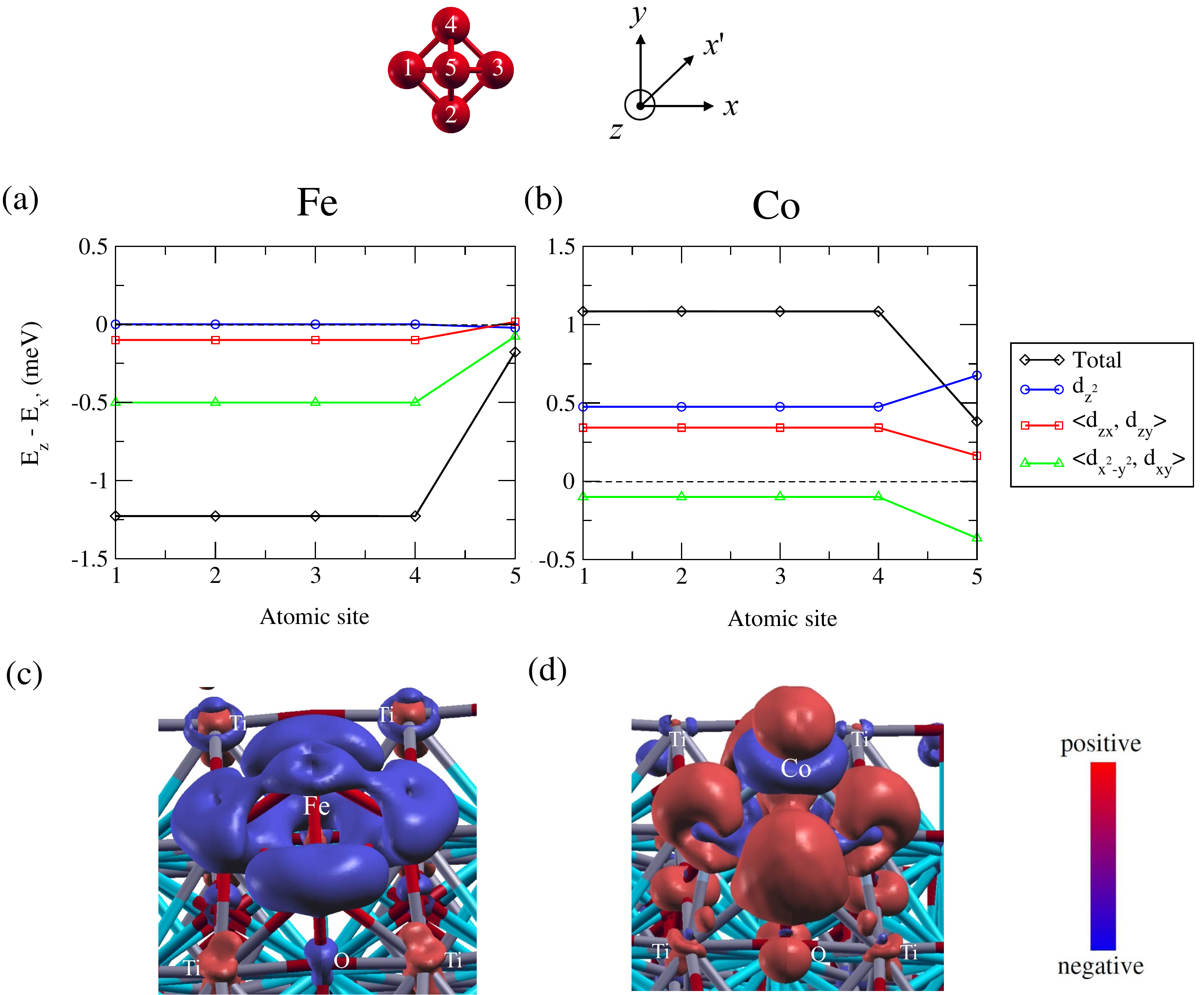}
\caption{Atom/$d$-orbitals-resolved MCA of Fe (a) and Co (b) clusters deposited on SrTiO$_3$. Due to symmetry, contributions from different orbitals in ($d_{zx}, d_{zy}$) and ($d_{x^2-y^2}, d_{xy}$) pairs are very similar so that their averaged values are presented for simplicity. Clear out-of-plane and in-plane MCA have been found for Fe and Co clusters, respectively. Real-space distribution of MCA for Fe (c) and Co (d) clusters is given. Note that red (blue) colors represent the regions favoring in-plane (out-of-plane) magnetization orientation. The MCA mainly from the base atoms for both clusters, and for Fe (Co) the MCA originates from $d$-orbitals of the cluster extending in-plane (out-of-plane).}
\label{pyramide_STO_mca}
\end{center}
\end{figure*}

The MCA is calculated by the formula
\begin{math}
\text{MCA}=E_{z}^{\text{band}}-E_{x'}^{\text{band}}
\end{math}
using as usual the magnetic force theorem. The MCA in the $xy$ plane is found to be extremely small. we have chosen the most symmetric in-plane direction $x'$ (see Fig. \ref{pyramide_STO_mca}) which has an azimuthal angle of $\phi$ = 45$^{\circ}$ with respect to $x$. Due to symmetry, this definition gives us almost similar contribution for each pair of ($d_{zx}, d_{zy}$) and of ($d_{x^2-y^2}, d_{xy}$) Fe(Co) orbitals, therefore, their averaged values are presented for the sake of simplicity.

In Fig.\ref{pyramide_STO_mca} (a) and (b) the local decomposition of MCA with different atomic sites as well as with different $d$-orbitals is presented for Fe$_5$$|$SrTiO$_3$ and Co$_5$$|$SrTiO$_3$, respectively. Note that only the contributions of clusters is shown. Interestingly, we find the opposite behavior of MCA for Fe and Co clusters deposited on SrTiO$_3$. The easy axis of magnetization is directed along out-of-plane for Fe cluster with a total MCA of $\sim$ $-$5.08 meV, on the contrary it is in-plane for Co  with a total MCA of $\sim$ 4.72 meV. For both elements, the atomically resolved MCA (black lines) reveals that the MCA is mainly dominated by the base atoms (numbered as 1 $\sim$ 4) and a relatively much smaller contribution from the top atom (numbered as 5). The value of MCA per atom is as large as $\sim$ $-$1.22 (1.08) meV/atom for base atom and $\sim$ $-$0.18 (0.38) meV/atom for the top atom of Fe(Co) cluster.

It is also interesting to note that the  MCA mainly originates from the $d$-orbitals of the cluster extending in-plane for Fe, namely ($d_{x^2-y^2}, d_{xy}$) orbitals, and out-of-plane for Co, namely, ($d_{z^2}, d_{zx}, d_{zy}$). 

Finally in Fig. \ref{pyramide_STO_mca} (c) and (d), we present the real-space distribution of MCA for Fe$_5$$|$SrTiO$_3$ and Co$_5$$|$SrTiO$_3$. The red colors represent in-plane magnetization direction, whereas the blue colors are out-of-plane easy axis. We can clearly se that the MCA mainly originates from the base atoms for both clusters, and for Fe(Co) the MCA originates from $d$-orbitals of the cluster extending in-plane (out-of-plane). In addition, due to hybridization between the states of TiO$_2$ surface and $d$-orbitals of the cluster, the Ti and O atoms close to the cluster gives a rather small contribution to MCA. For Fe, Ti atom slightly favors the in-plane easy axis and the easy axis of O atom is out-of-plane. In the case of Co, both Ti and O atoms around the cluster favor to in-plane magnetization direction.

As a consequence, we predict that the Fe$_5$ nanocrystals  should be magnetically stable and are thus good potential candidates for magnetic storage devices. 

\subsubsection{MCA analysis from perturbation theory}

Let us consider the perturbation of the total energy due to the spin-orbit coupling Hamiltonian $H_{\text{SO}}$ \cite{Bruno1989,Daalderop1994,Autes2006, Tsujikawa2009, Gimbert2012}. Since the first-order term vanishes  the second order perturbation term $\Delta E^{(2)}$ of the total energy has to be evaluated:
\begin{equation}
\Delta E^{(2)}=-\sum_{n \sigma \text{occ} \atop  n' \sigma' \text{unocc}} \frac{|\langle n \sigma | H_{\text{SO}}|n' \sigma'\rangle|^2}{E_{n' \sigma'}-E_{n  \sigma}}
\end{equation}
where $|n \sigma\rangle$ ($|n'\sigma'\rangle$) is an uperturbed occupied (unoccupied) state of energy $E_{n \sigma}$ ($E_{n' \sigma'}$) ,  $n$ denotes the index of the state and $\sigma$ its spin (which is still a good quantum number for the unperturbed state). Writing the eigenstates in an orthogonal basis of real atomic spin orbitals $\lambda \sigma$ centered at each atomic site $i$,   one can derive a rather cumbersome equation written explicitely in the Appendix of Ref. \onlinecite{Autes2006} (Eq. C.8). However it is possible to drastically simplifiy  Eq. C.8 by retaining only the diagonal terms of the density matrix which leads to the following expression:

\begin{equation}
\Delta E^{(2)}=A-\xi^2 \sum_{\lambda \mu}  |\langle \lambda \uparrow | H_{\text{SO}}| \mu \uparrow \rangle|^2 
\sum_{i \sigma \sigma'} \sigma \sigma' I_{i}(\lambda,\mu,\sigma,\sigma')
\end{equation}
where $A$ is a constant isotropic term and 
\begin{equation}
 I_{i}(\lambda,\mu,\sigma,\sigma')=\int_{-\infty}^{E_F}dE\int_{E_F}^{\infty} dE' \frac{n_{i\lambda \sigma}(E)n_{i\mu \sigma'}(E')}{E'-E}
\end{equation}
$n_{i\lambda \sigma}(E)$ ($n_{i\mu \sigma'}(E')$) being the projected density of states of occupied (unoccupied) states. The dominant terms  $I_{i}(\lambda,\mu,\sigma,\sigma')$ are the ones corresponding to a transition between an occupied  and an unoccupied state  presenting a high density of states below and above the Fermi level respectively.

The MCA defined as the difference of energy between the direction $z$ and $x$ can be decomposed in local atomic contributions MCA$_i$:

\begin{equation}
\text{ MCA}_i= \xi^2 \sum_{\lambda \sigma \atop \mu \sigma'} \sigma \sigma'  T_{\lambda,\mu}   I_{i}(\lambda,\mu,\sigma,\sigma')
\end{equation}

$ T_{\lambda,\mu}$ is the difference of the square of the spin-orbit matrix elements between two orientations of the magnetization $\mathbf M$:

\begin{equation}
T_{\lambda,\mu}=  |\langle\lambda\uparrow| L.S|\mu\uparrow\rangle|_{\mathbf{M}\parallel x}^2-  |\langle \lambda \uparrow | L.S| \mu \uparrow \rangle|_{\mathbf{M} \parallel z}^2
\end{equation}

Since  $ I_{i}(\lambda,\mu,\sigma,\sigma')$ is always positive, the sign of the matrix elements $\sigma \sigma' T_{\lambda,\mu}$ for a given transition between an occupied state $\lambda \sigma$ and an unocuppied state $\mu \sigma'$ will define the sign of the corresponding anisotropy. In practice there are a limited number of transitions and in addition spin-flip transitions are often negligible, therefore in most case $ \sigma \sigma'=1$. 

Let us now apply this perturbation expansion to the case of Iron and Cobalt clusters on SrTiO$_3$. First, it is clear from the PDOS analysis that the top atom will contribute negligibly to the total MCA. In contrast for both atoms the PDOS of the base atom shows that there are dominantly four occupied-unoccupied transitions that will dominate the MCA. Namely the transition $d_{x^2-y^2}\rightarrow d_{zx}$,  $d_{x^2-y^2}\rightarrow d_{xy}$,  $d_{zy}\rightarrow d_{zx}$, and $d_{zy}\rightarrow d_{xy}$ for Fe and  $d_{zx}\rightarrow d_{zx}$,  $d_{zx}\rightarrow d_{zy}$,  $d_{z^2}\rightarrow d_{zx}$, and $d_{z^2}\rightarrow d_{zy}$ for Co. From Eq. \ref{Eq.Tmatrix} it comes out that for Fe two transitions are large with a negative sign ($d_{x^2-y^2}\rightarrow d_{xy} \propto -4$, $d_{zy}\rightarrow d_{zx} \propto -1$ ) and two are small with a positive sign ($d_{x^2-y^2}\rightarrow d_{zx} \propto 1/2$, $d_{zy}\rightarrow d_{xy} \propto 1/2$ ). For Co we find the opposite trend: two transition have a large and positive sign ( $d_{z^2}\rightarrow d_{zx}\propto 3/2$, $d_{z^2}\rightarrow d_{zy}\propto 3/2$ ), one has a negative sign ( $d_{zx}\rightarrow d_{zy}\propto -1$ ) and the last one is diagonal and do not contribute ($d_{zx}\rightarrow d_{zx}=0$). Overall this shows that Fe pyramid favors out-of-plane magnetization while Co favors in plane magnetization. The main orbitals involved are $(d_{ xy},d_{x^2-y^2})$ for Fe and $(d_{zx},d_{zy},d_{z^2})$ for Co in agreement with the results presented in Sec. \ref{sec:local-MCA}

This type of analysis remains qualitative and applies preferentially to low dimensional systems presenting sharp features in their PDOS. 
Nevertheless, the arguments put forward are rather general and could be be very useful in the design of atomic-scale devices with optimized magnetic anisotropy. 
Note however, that if the nonsphericity of the Coulomb and exchange interaction\cite{Anisimov1993} starts to play a dominant role in the electronic structure of the system, then orbital polarization effects arise\cite{Desjonqueres2007,Desjonqueres2007a}  and our analysis of the MCA based on a pertubation treatment of the SOC only non longer applies, and more complex scenarii can occur as in the case of the giant magnetic anisotropy of single adatoms on MgO\cite{Ou2015}.

\section{ Conclusion }
\label{conclusion}

We investigated the electronic properties and MCA of Fe and Co slabs and nanoclusters interfaced with SrTiO$_3$ underlayer. Interestingly, a comparative study of Fe and Co freestanding slabs with their interface with SrTiO$_3$, revealed a tremendous impact of the latter on the MCA. Namely, the MCA contribution from the interfacial Fe layer in Fe$|$SrTiO$_3$ is quenched resulting in the loss of the perpendicular magnetic anisotropy (PMA) while for Co$|$SrTiO$_3$ , the anisotropy is changed from in-plane to the out-of-plane. This is explained by the orbital resolved analysis of hybridizations of Fe and Co $d$-orbitals of with those of Ti and $p_z$ orbital of O.

We also find a strong enhancement of out-of-plane and in-plane MCA for small Fe and Co clusters (containing only several atoms) upon  deposition on a SrTiO$_3$ substrate. The hybridization between the substrate and the $d$-orbitals of the cluster extending in-plane for Fe and out-of-plane for Co is at the origin of this enhancement of MCA. As a consequence, we predict that the Fe nanocrystals (even rather small) should be magnetically stable and are thus good potential candidates for magnetic storage applications.

\section{Acknowledgement}
The research leading to these results has received funding from the European Research Council under the European
Union’s Seventh Framework Programme (FP7/2007-2013)/ERC grant agreement No. 259297. This work was performed using HPC computation resources from GENCI-[TGCC] (Grant No. 2015097416).
\\

\appendix

\section{Expression of the $\mathbf{T}$ matrix}
The matrix elements of the spin-orbit coupling Hamiltonian in the $d$ orbital basis (ordered as $d_{xy}, d_{zy},  d_{zx}, d_{x^2-y^2}, d_{z^2}$) are written explictely in Apppendix A of Ref. \onlinecite{Autes2006} for an arbitrary orientation of the magnetization  defined by the altitude angle and the azimuth angle $(\theta,\phi)$ .
If we define the MCA as the total energy difference between a magnetization along $z$ ($\theta=\phi=0$) and a magnetization along an arbitrary direction $\mathbf{n} (\theta,\phi)$ the corresponding $\mathbf{T}$ matrix reads:

\begin{scriptsize}
\begin{equation}
\frac{1}{4}
\begin{bmatrix} 
     0                                        &  \sin^2\theta \sin^2\phi    & \sin^2\theta \cos^2\phi       & -4 \sin^2\theta                  & 0                                       \\
    \sin^2\theta \sin^2\phi     &  0                                        & - \sin^2\theta                       &  \sin^2\theta \cos^2\phi   & 3\sin^2\theta \cos^2\phi \\ 
     \sin^2\theta \cos^2\phi   & - \sin^2\theta                     &  0                                          & \sin^2\theta \sin^2\phi     &  3\sin^2\theta \sin^2\phi   \\
   -4 \sin^2\theta                   &   \sin^2\theta \cos^2\phi  &  \sin^2\theta \sin^2\phi      &   0                                        & 0                                            \\
     0                                        &  3\sin^2\theta \cos^2\phi &  3\sin^2\theta \sin^2\phi   & 0                                           & 0                                            \\
\end{bmatrix}
\end{equation}
\end{scriptsize}

 If $\mathbf{n}$ is along $x$ ($\theta=\pi/2, \phi=0$), $\mathbf{T}$ takes the form:

\begin{equation}
\frac{1}{4}
\begin{bmatrix} 
     0 &  0 & 1 & -4  & 0 \\
    0 &  0 & -1 & 1  & 3 \\
    1 & -1 &  0 & 0  & 0 \\
   -4 &  1 &  0 & 0  & 0 \\
    0 &  3 &  0 & 0  & 0 \\
\end{bmatrix}
\end{equation}

While if instead of $x$ we take the more symmetric in plane $x'$ direction ( ($\theta=\pi/2, \phi=\pi/4$) we find for  $\mathbf{T}$:
\begin{equation}
\label{Eq.Tmatrix}
\frac{1}{4}
\begin{bmatrix}
     0   &  1/2 & 1/2 & -4  & 0 \\
    1/2 &  0 & -1 & 1/2  & 3/2 \\
    1/2 & -1 &  0 & 1/2  & 3/2 \\
   -4 &  1/2 &  1/2 & 0  & 0 \\
    0 &  3/2 &  3/2 & 0  & 0 \\
\end{bmatrix}
\end{equation}

A positive sign means an easy axis along $\mathbf{n}$  and a negative sign an easy axis along $z$.

\bibliographystyle{apsrev}
\bibliography{References}

\end{document}